\documentclass[aoas,MSNbibl,nameyear,dvips]{arximspdf}
\usepackage{mathbh,dcolumn}
\usepackage{graphicx}

\doi{10.1214/13-AOAS711} 
\volume{8}
\issue{1}
\pubyear{2014}
\firstpage{622}
\lastpage{647}

\makeatletter
\newcolumntype{d}[1]{D{.}{.}{#1}}
\newcommand{\E}{\mathrm{E}}
\def\R{{\mathbb{R}}}
\def\P{{\mathbb{P}}}

\newcommand{\GEV}{\mathrm{GEV}}
\newcommand{\GPD}{\mathrm{GPD}}
\makeatother

\begin{document}
\begin{frontmatter}

\title{Modeling extreme values of processes observed at irregular time
steps: Application to significant~wave height}
\runtitle{Modeling extreme values of processes}

\pdftitle{Modeling extreme values of processes observed at irregular time
steps: Application to significant~wave height}

\begin{aug}
\author[A]{\fnms{Nicolas} \snm{Raillard}\corref{}\ead[label=e1]{nicolas.raillard@gmail.com}\thanksref{m1,m2,m3}},
\author[B]{\fnms{Pierre} \snm{Ailliot}\thanksref{m1}}
\and
\author[C]{\fnms{Jianfeng} \snm{Yao}\thanksref{m4}}
\runauthor{N. Raillard, P. Ailliot and J. Yao}
\affiliation{Universit\'{e} de Brest\thanksmark{m1},
IFREMER\thanksmark{m2}, Universit\'{e} de Rennes 1\thanksmark{m3} and\\ University of
Hong Kong\thanksmark{m4}}
\address[A]{N. Raillard\\
Laboratoire de Math\'{e}matiques\\
\quad de Bretagne Atlantique\\
UMR 6205\\
Universit\'{e} de Brest\\
France\\
Laboratoire d'Oc\'{e}anographie Spatiale\\
IFREMER\\
France\\
and\\
Insititut de Recherche Math\'{e}matique de Rennes\\
UMR 6625\\
Universit\'{e} de Rennes 1\\
France\\
\printead{e1}} 
\address[B]{P. Ailliot\\
Laboratoire de Math\'{e}matiques\\
\quad de Bretagne Atlantique\\
UMR 6205\\
Universit\'{e} de Brest\\
France}
\address[C]{J. Yao\\
Department of Statistics\\
\quad and Actuarial Sciences\\
University of Hong Kong\\
Hong Kong}
\end{aug}

\received{\smonth{2} \syear{2013}}
\revised{\smonth{12} \syear{2013}}

%
\begin{abstract}
This work is motivated by the analysis of the extremal behavior of buoy
and satellite data describing wave conditions in the North Atlantic
Ocean. The available data sets consist of time series of significant
wave height (Hs) with irregular time sampling. In such a situation, the
usual statistical methods for analyzing extreme values cannot be used
directly. The method proposed in this paper is an extension of the
peaks over threshold (POT) method, where the distribution of a process
above a high threshold is approximated by a max-stable process whose
parameters are estimated by maximizing a composite likelihood function.
The efficiency of the proposed method is assessed on an extensive set
of simulated data. It is shown, in particular, that the method is able
to describe the extremal behavior of several common time series models
with regular or irregular time sampling. The method is then used to
analyze Hs data in the North Atlantic Ocean. The results indicate that
it is possible to derive realistic estimates of the extremal properties
of Hs from satellite data, despite its complex space--time sampling.
\end{abstract}

%

\pdfkeywords{Extreme values,
time series,
max-stable process,
composite likelihood,
irregular time sampling,
significant wave height,
satellite data}

\end{frontmatter}

\section{Introduction}\label{sec1}
\begin{figure}

\includegraphics{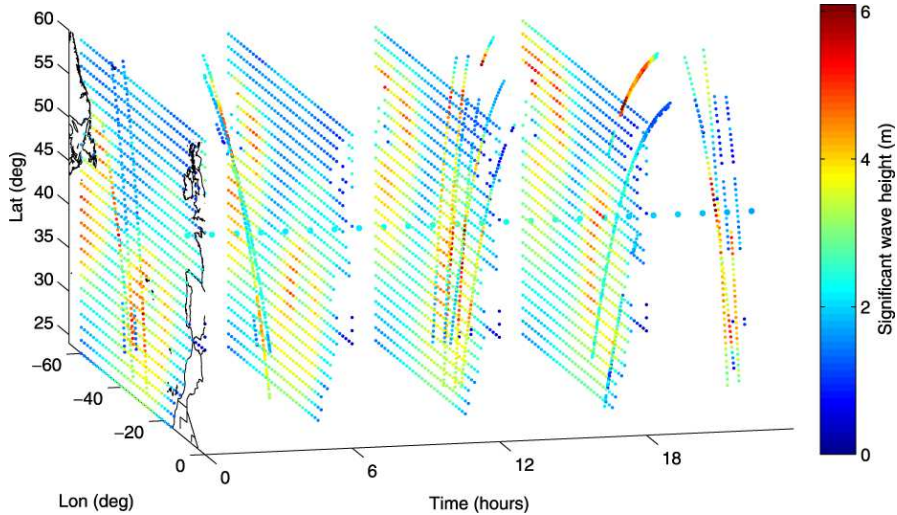}

\caption{3D representation of the available data
for 12/16/2002. The 2D fields at times 0, 6, 12 and 18 correspond to
reanalysis data, the 24 observations at the same location correspond to
buoy data (hourly data), and the other data to the various satellite
tracks.}\label{figdatadispo}
\end{figure}

Extreme events are a major concern in statistical modeling, and
appropriate methods are needed to derive estimates of the extremal
properties of various phenomena from complex observations. For example,
the 100-year return level of significant wave height is often used in
the design of marine structures as a criterion to characterize the
extreme waves that a structure may face during its lifetime.
Significant wave height, generally denoted Hs, can be interpreted as a
measure of an average wave height in a sea state. Three main sources of
data can be used to assess the statistical properties of Hs:
\begin{itemize}
\item\textit{Reanalysis data} provide long time series (typically a
few decades) everywhere in the oceans at a regular time step and
without missing values, but tend to smooth out extreme values.
\item\textit{Buoy data} are generally more accurate, but cover shorter
time periods (typically a few years with missing values) and have a
poor spatial distribution.
\item\textit{Satellite data} are also accurate observations and are
available on the last 20~years. However, the time series obtained by
selecting all satellite data available at a given location exhibits
complex irregular time sampling depending on the number and tracks of
the operating satellites.
\end{itemize}
A typical example of the data coverage over a 24-hour time window in
the North Atlantic is given in Figure~\ref{figdatadispo}. The
motivation for the work presented here was to develop statistical
methods for analyzing the extremal properties of Hs based on such data
sets. In particular, our proposed method can be used to estimate
various characteristics of the extremal behavior of processes (e.g.,
high quantiles, return periods, storm durations, etc.) observed at
regular or irregular time steps, whereas most existing methods are
inappropriate in the latter case.

%
%

Two methods are commonly used for estimating extreme quantiles [see,
e.g., \citet{Embrechts1997}, \citet{Coles2001}, \citet
{Beirlant2004}, \citet{deHaan2006a}
and the references therein]. The first one, generally referred as the
block maxima method, relies on probabilistic results, which suggest the
use of the generalized extreme-value ($\GEV$) distribution for modeling
the maximum of a large number of identically distributed random
variables. The main drawback of this approach is the waste of data
induced by taking the maximum over a large block, typically one year
for meteorological applications, before fitting the $\GEV$
distribution. The second approach, generally referred to as peaks over
threshold (POT), involves keeping only the observations above a certain
threshold chosen to be sufficiently high to ensure that the
distribution of excesses above that threshold is well approximated by a
generalized Pareto distribution ($\GPD$). A problem that arises in
using the POT approach for time series is that clusters of consecutive
dependent exceedances are generally observed, particularly when the
time-lag between successive observations is smaller than the
characteristic duration of extreme events. In this case a
``declustering'' step is generally applied before fitting the $\GPD$
distribution to the sample of cluster maxima.
This method also leads to a waste of data since only one value per
cluster is kept to fit the $\GPD$, and it relies on arbitrary rules for
declustering the data. Those rules are even more difficult to choose in
the presence of missing values or irregular time sampling.

Another approach, initially proposed in \citet{Smith1997}, retains all
of the exceedances and models them as a first-order Markov chain whose
transition kernel is derived from bivariate extreme value theory. A
concurrent strategy, adopted, for example, in \citet{bortot2011latent},
consists of modeling the exceedances conditionally on a latent process.
Both approaches have been successfully applied to various
meteorological time series [see, e.g., \citet{Ribatet2009}], but they
cannot be adapted straightforwardly to time series with irregular
sampling [see also \citet{Reich2013}]. In this paper, we propose an
alternative approach in which the time series of exceedances above a
high threshold is assumed to be a realization of a censored max-stable
process. This model is motivated by probabilistic results from extreme
value theory. Since the full likelihood cannot be obtained
analytically, we follow recent works on spatial and space--time
extremes, and base the statistical inference on a composite likelihood
approach. The proposed model can be easily simulated, thereby allowing
estimation of various quantities of interest for applications, such as
quantiles, return periods or the characteristics of sojourns above high
thresholds using Monte Carlo simulations. Parametric bootstrap can also
be used to assess the accuracy of the estimators.

The model is introduced in Section~\ref{secmodel}. Section~\ref{secestim} studies the composite likelihood approach and discusses the
properties of the estimators using simulations. Section~\ref{secsim}
describes simulation results to validate our approach on classical
time-series models and Section~\ref{secappli} discusses its
application to Hs data.\eject

\section{Censored max-stable processes}\label{secmodel}


\subsection{Threshold models and censoring in the independent case}\label{seciid}
Probably the most classical approach to modeling the extremal
properties of an independent and identically distributed (i.i.d.) sample
$X_1,\ldots,X_n$ consists in using ``block maxima.'' It relies on
probabilistic results originating in \citet{Fisher1928} which suggest
approximating the distribution of $\textstyle M_n = \max_{i = 1,\ldots, n} X_i$ by a $\GEV$ distribution with the following cumulative
distribution function (c.d.f.):
%
%
\begin{equation}
\label{GEV} F(x;\mu,\sigma,\xi) = \cases{ \displaystyle\exp\biggl\{-
\biggl[1+ \xi\frac{x-\mu}{\sigma} \biggr]^{-1/\xi} \biggr\}&\quad$(\xi\neq0)$,
\vspace*{5pt}\cr
\displaystyle\exp\biggl\{-\exp\biggl[-\frac{x-\mu}{\sigma} \biggr]
\biggr\} &\quad$( \xi= 0)$,}
\end{equation}
defined for $x$ such that $1+\xi\frac{x-\mu}{\sigma}>0$ with
parameters $\mu\in\R$, $ \sigma>0$ and $\xi\in\R$. For
applications, the data are grouped into blocks of equal length, and a
$\GEV$ distribution is fitted to the sample of block maxima. The choice
of block size is critical in practice. For environmental time series,
the $\GEV$ distribution is generally fitted to the time series of
annual maxima to remove seasonal effects. This leads to a waste of
data, and the sample size used to estimate the three parameters of the
$\GEV$ distribution is the number of years with no or few missing
values (a few decades in the best-case scenario). Although many methods
have been proposed to provide estimates which exhibit a good behavior
on small samples [see, e.g., \citet{Ailliot2008} and the references
therein], such estimation remains an important issue in applications.

The POT approach is the classical alternative to the block maxima
approach. It is less wasteful of data since it keeps all data above a
high threshold $u$, which is chosen such that the conditional
distribution $\P[X_i \leq x | X_i>u]$ is well approximated by a $\GPD$
with c.d.f.
\[
G(x;\mu,\sigma,\xi) = \cases{ \displaystyle1- \biggl( 1 + \xi\frac
{x-\mu}{\sigma}
\biggr)^{-1/\xi}&\quad$(\xi\neq0)$,
\vspace*{5pt}\cr
\displaystyle 1- \exp\biggl[-\frac{x-\mu}{\sigma}
\biggr] &\quad$(\xi= 0)$,}
\]
defined for $x \geq\mu$ such that $1 + \xi\frac{x-\mu}{\sigma}
\geq
0$ with $\mu=u$ and parameters $\sigma>0$ and $\xi\in\R$. Again, the
use of the $\GPD$ is motivated by probabilistic results, and various
methods have been proposed for estimating parameters, based on the
sample of exceedances and the choice of $u$, although the latter is a
more difficult problem [see \citet{Davison1990}]. Once $u$ is chosen,
the standard method for estimating the unknown parameters is to
maximize the likelihood function
%
%
\begin{eqnarray}\label{eqPOTLL2}
&& L(\lambda, \sigma, \xi;X_1,\ldots,X_n)\nonumber
\\
&&\qquad = \lambda^{N_u} (1-\lambda)^{n-N_u} \prod
_{i \in\{1,\ldots,n\}|X_i> u} g(x_i;u,\sigma, \xi)
\\
&&\qquad = \prod_{i \in\{1,\ldots,n\}|X_i\leq u} \lambda\prod
_{i \in\{1,\ldots,n\}
|X_i> u} (1-\lambda) g(x_i;u,\sigma, \xi),\nonumber
\end{eqnarray}
where $\lambda=\P(X_i \leq u)$, $N_u$ is the number of observations
below the threshold $u$, and $g(x;\mu,\sigma, \xi)$ is the probability
density function (p.d.f.) of the $\GPD$.
It is well known that the conditional distribution of the exceedances
of a $\GPD$ above an arbitrary threshold is also a $\GPD$, which allows
us to interpret (\ref{eqPOTLL2}) as the likelihood of an i.i.d. sample of
a $\GPD$ censored at the threshold $u$.
More precisely, let $\tilde X_1,\tilde X_2,\ldots,\tilde X_n$ be an i.i.d.
sample of a $\GPD$ with parameter $(\mu,\tilde\sigma,\xi)$, and
consider the censored random variable
\[
Y_i=u\mathbh{1}_{[\tilde X_i \leq u]}+\tilde X_i
\mathbh{1}_{[\tilde
X_i > u]}= \cases{ u, &\quad if $\tilde X_i \leq u$,
\vspace*{2pt}\cr
\tilde X_i, &\quad if $\tilde X_i > u$,}
\]
where the threshold $u$ belongs to the support of the GPD distribution.
We have $P(Y_i=u)=\lambda$ with $\lambda=G(u;\mu,\tilde\sigma,\xi)$
and, for $x > u$,
%
\begin{eqnarray}
\nonumber
\P(Y_i \geq x)&=&\P(\tilde X_i \geq u)\P(
\tilde X_i \geq x|\tilde X_i \geq u)
\\
\nonumber
&=& \bigl(1-G(u;\mu,\tilde\sigma,\xi) \bigr) \frac{1-G(x;\mu,\tilde
\sigma,\xi
)}{1-G(u;\mu,\tilde\sigma,\xi)}
\\
\nonumber
&=&(1-\lambda) \bigl(1-G(x;u,\sigma,\xi) \bigr)
\end{eqnarray}
with $\sigma=\tilde\sigma(1+\xi\frac{u-\mu}{\tilde\sigma
}
)$, and, thus, (\ref{eqPOTLL2}) is the likelihood of $
(Y_1,\ldots,Y_n )$. Finally, the assumptions made when using the POT
approach are equivalent to assuming that the original sample
$X_1,\ldots,X_n$ satisfies
%
%
\begin{equation}
\label{eqmodiid} u\mathbh{1}_{[X_i \leq u]}+X_i\mathbh{1}_{[X_i > u]}=u
\mathbh{1}_{[\tilde X_i \leq
u]}+\tilde X_i\mathbh{1}_{[\tilde X_i > u]}
\end{equation}
for all $i \in\{1,\ldots,n\}$, where $\tilde X_1,\ldots,\tilde X_n$
is an
i.i.d. sample of a $\GPD$.

We will see below that this interpretation of the POT approach in terms
of censoring has advantages for modeling purposes. From a numerical
point of view, it can be viewed as a reparametrization of the
likelihood function. Maximizing (\ref{eqPOTLL2}) over $(\lambda,
\sigma, \xi)$ leads to the estimate $\hat{\lambda}=\frac{N_u}{n}$ for
$\lambda$, which has the desirable properties of being easy to
interpret and of being independent of the estimates of $\sigma$~and~$\xi
$. At the same time, maximizing (\ref{eqPOTLL2}) over $(\mu,\tilde
\sigma,\xi)$ leads to a more complicated three-dimensional optimization
problem and correlated estimates, which can be problematic for certain
applications, as discussed in Section~\ref{secappli} and in \citet
{Ribereau2011}.

Although the $\GPD$ distribution is the most common choice for modeling
exceedances over a high threshold, other tail approximations have been
proposed in the literature. In particular, it is known that GEV and GPD
have the same tail behavior [see \citet{Drees06}], and this suggests
that similar results will be obtained if we model the distribution of
$\tilde X_i$ by a $\GEV$ distribution instead of a $\GPD$. Various
tests on simulated samples have confirmed that both approximations lead
to similar results in practice.

The tail approximations discussed above remain valid for dependent
sequences under mild conditions [see \citet{Leadbetter1983}], which
justifies the use of both the block maxima and POT approaches for
analyzing the extremes of time series. One difficulty with using the
POT approach in this context is that clusters of consecutive dependent
exceedances are generally observed, whereas the likelihood function
(\ref{eqPOTLL2}) is the joint distribution of the exceedances only if
they are independent. The most common POT method thus includes a
declustering step, with the maxima within each cluster kept only to
obtain a sample of approximately independent exceedances. It also leads
to waste data and thus degrades the quality of the estimates.
Alternative strategies, which keep all the exceedances in the fitting
procedure but correct the estimation of the uncertainty of the
estimators to account for dependence, are proposed in \citeauthor{Fawcett2007}
(\citeyear{Fawcett2007,fawcett2012estimating}).

\subsection{Censored max-stable processes}\label{secdep}

We now consider a sample $X_{t_1},\ldots,X_{t_n}$ of a stochastic process
$\{X_t\}$ observed at times $t_1,\ldots,t_n$. It is generally assumed in
the literature that the observations are available at a regular time
step [i.e., $t_{i+1}-t_i=t_{j+1}-t_j$ for all $(i,j) \in\{1,\ldots,n-1\}
$], but we are interested in a method that is sufficiently flexible to
deal with irregular time sampling. We thus propose analyzing the
extremal behavior of such a data set by extending the POT approach and
modeling the exceedances of the process $\{X_t\}$ above a high
threshold $u$ as a censored max-stable process. The theory of
max-stable processes [see \citet{Haan1984}, \citet
{deHaan2006a}] is a natural
generalization of the traditional univariate max-stable theory used to
motivate the choice of the $\GEV$ distribution in the i.i.d. case. Several
families of max-stable processes have been proposed in the literature
[see, e.g., \citet{Smith1990}, \citet{Schlather2002}]. Our
focus here is on the
specific \textit{Gaussian extreme value process} introduced in \citet
{Smith1990}, although the methodology introduced herein can easily be
adapted to other models.
More precisely, we assume that
%
%
\begin{equation}
\label{eqmod} u\mathbh{1}_{[X_t \leq u]}+X_t\mathbh{1}_{[X_t > u]}=u
\mathbh{1}_{[\tilde X_t \leq
u]}+\tilde X_t\mathbh{1}_{[\tilde X_t > u]}
\end{equation}
holds for all $t$, where $u$ is a fixed threshold and $\{\tilde X _t\}$
is a stationary Gaussian extreme value process with parameter $\theta
= (\mu,\sigma,\xi,\nu) \in(-\infty,+\infty) \times
(0,+\infty) \times(-\infty,+\infty) \times(0,+\infty)$, as
defined below:
\begin{itemize}
\item The marginal distribution of $\{\tilde X _t\}$ is a $\GEV$
distribution with parameter $(\mu,\sigma,\xi)$. This assumption implies
that the process $\{Z_t\}$ obtained by applying the following marginal
transformation,
%
%
\begin{equation}
Z_t=-\frac{1}{\log(F(\tilde{X}_t;\mu,\sigma,\xi))} \label{eqtranmarg},
\end{equation}
is a stationary process with a unit Fr\'echet marginal distribution
[i.e., $\GEV$ distribution with parameter $(1,1,1)$].
\item We further assume that
%
%
\begin{equation}
Z_t= \max\biggl\{ \frac{\zeta_i}{\nu\sqrt{2 \pi}} \exp\biggl(-\frac
{(s_i- t)^2}{2\nu^2}
\biggr) \biggr\}, \label{eqdefGEVP}
\end{equation}
where $\{(\zeta_i, s_i), i \geq1\}$ denote the points of a Poisson
process on $(0,\infty) \times\R$ with intensity measure $\zeta
^{-2}\,d\zeta\times ds$.
\end{itemize}

We focus on Gaussian extreme value processes because they have a nice
meteorological interpretation [see \citet{Smith1990}], can easily handle
observations available at irregular time steps, and are quick and easy
to simulate [see \citet{Schlather2002}]. The following sections also
show that these processes provide a flexible class of models which is
able to describe the extremal behavior of most standard time series
models and the Hs data considered in this work. The parameters $\mu$,
$\sigma$~and~$\xi$ are related to the marginal distribution and can be
interpreted, respectively, as location, scale and shape parameters,
whereas the parameter $\nu$ is related to the temporal structure of the
process and may be interpreted as the typical duration of storms. More
precisely, we have [see \citet{Smith1990}]
%
%
\begin{equation}
\label{bvZ} \P(Z_{t_1} \leq z_{t_1}, Z_{t_2} \leq
z_{t_2}) = F_Z(z_{t_1},z_{t_2};\nu) =
\exp\bigl[-V(z_{t_1},z_{t_2};\nu) \bigr],
\end{equation}
where
%
%
\begin{equation}
\label{defVfrech} V(z_{t_1},z_{t_2};\nu) = \frac{1}{z_{t_1}} \Phi
\biggl( \frac{a}{2} + \frac{1}{a}\log\frac{z_{t_2}}{z_{t_1}} \biggr) +
\frac{1}{z_{t_2}} \Phi\biggl( \frac{a}{2} + \frac{1}{a}\log
\frac{z_{t_1}}{z_{t_2}} \biggr)
\end{equation}
with $a=\frac{|t_1-t_2|}{\nu}$, and $\Phi$ is the c.d.f. of the standard
normal distribution. The limit cases $\nu\rightarrow0$ and $\nu
\rightarrow+\infty$ correspond to independence and perfect dependence,
respectively.

Applying the inverse marginal transformation leads to the following
bivariate c.d.f. for the Gaussian extreme value process $\{\tilde X_t\}$:
%
%
\begin{eqnarray}\label{bvX}
\qquad F_{\tilde X}(\tilde x_{t_1}, \tilde x_{t_2};
\theta)&=&\P(\tilde X_{t_1} \leq\tilde x_{t_1}, \tilde
X_{t_1} \leq\tilde x_{t_2})
\nonumber\\[-8pt]\\[-8pt]
&=& \exp\biggl[-\frac{1}{z_{t_1}} \Phi\biggl( \frac{a}{2}
+ \frac
{1}{a}\log\frac{z_{t_2}}{z_{t_1}} \biggr) -\frac{1}{z_{t_2}} \Phi
\biggl( \frac
{a}{2} + \frac{1}{a}\log\frac{z_{t_1}}{z_{t_2}} \biggr)
\biggr]\nonumber
\end{eqnarray}
with $z_{t_i}=\frac{-1}{\log F(\tilde x_{t_i};\mu,\sigma,\xi)}$.

The\vspace*{1.5pt} bivariate distribution of $ (Y_{t_1},Y_{t_2} )$, where
$Y_{t}=u\mathbh{1}_{[\tilde X_t \leq u]}+\tilde X_t\mathbh
{1}_{[\tilde X_t > u]}$
is the censored Gaussian extreme value process, has the following
bivariate p.d.f.:
%
%
\begin{equation}
\label{CPLCSP} p_Y(y_{t_1},y_{t_2};\theta) =
\cases{
\displaystyle F_{\tilde X}(u,u;\theta), &\quad if  $y_{t_1} = u$ and
$y_{t_2} = u$,
\vspace*{5pt}\cr
\displaystyle \frac{\partial F_{\tilde X}}{\partial\tilde x_{t_1}}(y_{t_1},u;\theta
), & \quad if $y_{t_1} > u$ and $y_{t_2} = u$,
\vspace*{5pt}\cr
\displaystyle \frac{\partial F_{\tilde X}}{\partial\tilde x_{t_2}}
(u,y_{t_2};\theta), &\quad if $y_{t_1} = u$ and
$y_{t_2} > u$,
\vspace*{5pt}\cr
\displaystyle \frac{\partial^2 F_{\tilde X}}{\partial\tilde x_{t_1} \partial
\tilde x_{t_2} }(y_{t_1},y_{t_2};
\theta), &\quad if $y_{t_1} >u$ and $y_{t_2}>u$}
\end{equation}
with respect to the product measure $m \otimes m$, where $m(dx) =
\delta
_u(dx) + dx$ is the measure obtained by mixing the Dirac measure at $u$
with the Lebesgue measure.

Similar approximations, motivated by probabilistic results from
bivariate extreme value theory, are used in \citet{Smith1997} and
\citet{Ribatet2009} to model the bivariate distribution of successive
exceedances. Those papers further assume that the censored process is a
Markov process and the full likelihood function is derived from the
bivariate distributions. More recently, threshold versions of
max-stable processes have also been proposed in a space--time context
[see \citet{huser2012space}, \citet{jeon2012dependence}].

\section{Parameter estimation}\label{secestim}

\subsection{Composite likelihood estimators}\label{secCL}
In this section $(y_{t_1},\ldots,y_{t_n})\in\break (u,+\infty)^n$ denotes a
realization of a Gaussian extreme value process $\{Y_t\}$ with\vadjust{\goodbreak} unknown
parameter $\theta^* = (\mu^*,\sigma^*,\xi^*,\nu^*)$ censored at the
threshold $u\geq0$ and observed at times $t_1,\ldots,t_n$. There is no
known tractable expression for the full likelihood. However, since the
marginal and bivariate distributions have tractable expressions,
statistical inference can be based on one of the composite likelihood
functions introduced below [see also \citet{Lindsay1988},
\citet{VV2005}, \citet{Varin2008}, \citet{Cox2004},
\citet{Padoan2009}]:
\begin{itemize}
\item
The \textit{independent likelihood} function is defined as
%
%
\begin{equation}
\label{IL} \mathrm{IL}(\theta;y_{t_1},\ldots,y_{t_n})=\prod
_{i=1}^{n} p_Y(y_{t_i};
\theta),
\end{equation}
where $p_Y(y_{t};\theta)$ is the p.d.f. of the marginal distribution of
$Y_{t}$ with respect to the measure $m$. It is given by
\[
p_Y(y_{t};\theta) = \cases{ F(u;\mu,\sigma,\xi), &\quad if $y_{t} = u$,
\vspace*{2pt}\cr
f(y_{t};\mu,\sigma,\xi), &\quad if $y_{t} > u$,}
\]
where $F$ and $f$ denote the c.d.f. and the p.d.f. of the $\GEV$
distribution, respectively. It corresponds to the likelihood function
of an i.i.d. sample of a censored $\GEV$ distribution (see Section~\ref{seciid}) and does not invoke the parameter $\nu$, which describes the
dependence structure of the process. We denote by $\mathrm{MILE}$ the
estimator of $(\mu,\sigma,\xi)$ obtained by maximizing this function.
\item The \textit{pairwise likelihood} function is defined as
%
%
\begin{equation}
\label{CL} \mathrm{PL}(\theta;y_{t_1},\ldots,y_{t_n})=\prod
_{i=1}^{n-1} \prod
_{j>i} p_Y(y_{t_i},y_{t_j};
\theta)^{\omega_{t_i,t_j}}
\end{equation}
with $p_Y(y_{t_i},y_{t_j};\theta)$ given by (\ref{CPLCSP}) and
$\omega
_{t_i,t_j} \in\{0,1\}$ indicating whether the pair of observations
$(y_{t_i},y_{t_j})$ contributes to the pairwise likelihood function.
This approach has already been considered for time series with regular
time sampling [see \citet{Varin2008} and the references therein].
It is
generally assumed that
%
%
\begin{equation}
\label{eqstrag1} \omega_{t_i,t_j} = \mathbh{1}_{[|i-j|\leq K]},
\end{equation}
such that only the pairs of observations less than $K$ time units apart
are retained to build the pairwise likelihood function. Hereafter,
$\mathrm{PL}_K$ denotes the corresponding pairwise likelihood function and
$\mathrm{MPL}_K\E$ the estimator obtained by maximizing this function.
Retaining only the neighboring observations (i.e., using $K=1$) has
clear computational benefits since it significantly reduces the number
of terms in the product (\ref{CL}). It may also lead to more efficient
estimators in practice [see \citet{Varin2008} and Section~\ref{secsimu}]. Another strategy is to retain the pairs of observations
separated by a time lag smaller than $K$ and take
%
%
\begin{equation}
\label{eqstrag2} \omega_{t_i,t_j} = \mathbh{1}_{[|t_i-t_j|\leq K]}.
\end{equation}
This second strategy is similar to the first one when the process is
observed at regular time sampling but differs in the irregular case,
which will be further discussed using simulations in Section~\ref{secsimu}.
\item The \textit{Markovian likelihood} function is defined as
%
%
\begin{eqnarray}\label{ML}
\mathrm{ML}(\theta;y_{t_1},\ldots,y_{t_n})&=&p_Y(y_{t_1};
\theta) \prod_{i=2}^{n}p_Y(y_{t_i}
| y_{t_{i-1}};\theta)\nonumber
\\
&=& \frac{\prod_{i=2}^{n}p_Y(y_{t_i}, y_{t_{i-1}};\theta)}{\prod
_{i=2}^{n-1}p_Y(y_{t_{i}};\theta)}
\\
&=& \frac{\mathrm{PL}_1(\theta;y_{t_1},\ldots,y_{t_n})}{\mathrm{IL}(\theta;y_{t_2},\ldots,y_{t_{n-1}})}\nonumber
\end{eqnarray}
and $\mathrm{MMLE}$ denotes the estimator obtained by maximizing this
function. This estimator is considered for comparison purposes. Indeed,
when the process is observed at a regular time step, we retrieve the
Markovian model considered in \citet{Smith1997} and \citet
{Ribatet2009}
for the specific bivariate max-stable distribution associated with the
Gaussian extreme value process.
\end{itemize}

From a numerical point of view, we find it useful to use a two-stage
procedure, where the parameters $(\mu,\sigma,\xi)$ of the marginal
distribution are first estimated by maximizing the independent
likelihood function, and the dependence parameter $\nu$ is then
estimated by maximizing the pairwise likelihood function in $\nu$ with
the parameters of the marginal distribution kept fixed at the values
obtained in the first step. Doing so permits to reduce the
computational time (the independent likelihood function can be
evaluated quickly compared to the pairwise likelihood function) and
avoid the divergence problems which may occur when optimizing the
pairwise likelihood function simultaneously over all parameters with an
inappropriate starting point. Then, we can perform a global
optimization of the pairwise likelihood function over the four
parameters with the estimates obtained after the two-stage procedure
being used as the starting point. We performed various numerical
experiments, which confirmed that the estimators obtained from the
two-stage procedure are suboptimal compared to full optimization [see
\citet{dos2008copula} for a discussion in the context of copulas].

The asymptotic properties of the composite likelihood estimators for
max-stable processes have been studied in several recent papers. In
\citet{Padoan2009} it is shown that they are consistent and
asymptotically normal when the sample consists of i.i.d. replicates of
max-stale fields. Two recent papers [\citet{huser2012space},
\citet{jeon2012dependence}] consider the general case of
temporally and
spatially dependent max-stable processes. Although the asymptotic
results developed in \citet{jeon2012dependence} apply to the estimators
considered in this paper, for the sake of completeness, an alternative
proof of the consistency of the $\mathrm{MPL}_1\E$ is provided in
the supplementary article \citet{Raillard2013}. That proof is provided
in an idealized situation with no censoring and known marginal
distributions. Moreover, since the asymptotic covariance of the
estimator is too complicated to compute for practical applications, we
use parametric bootstrap [see, e.g., \citet{Benton2002}] to approximate
the distribution of the estimators and provide confidence intervals, as
discussed in greater detail in Section~\ref{secappli}.

\subsection{Simulation study}\label{secsimu}

A simulation study was undertaken to assess the accuracy of the
estimators introduced in Section~\ref{secCL}. Random samples were
generated from a Gaussian extreme value process with parameters $\mu
=0$, $\sigma=1$ and $\xi=0.3$, which correspond to realistic values for
environmental applications. We performed various experiments to
investigate how the accuracy of the estimators is affected by the
sample size $n$, the dependence parameter $\nu$, the threshold $u$ and,
finally, by the strategy used to select the pairs of observations that
contribute to the pairwise likelihood function. For each experiment,
$1000$ independent sequences of a Gaussian extreme value process were
generated and the various estimators of $(\mu,\sigma,\xi,\nu)$ computed
for each sequence.

\begin{figure}

\includegraphics{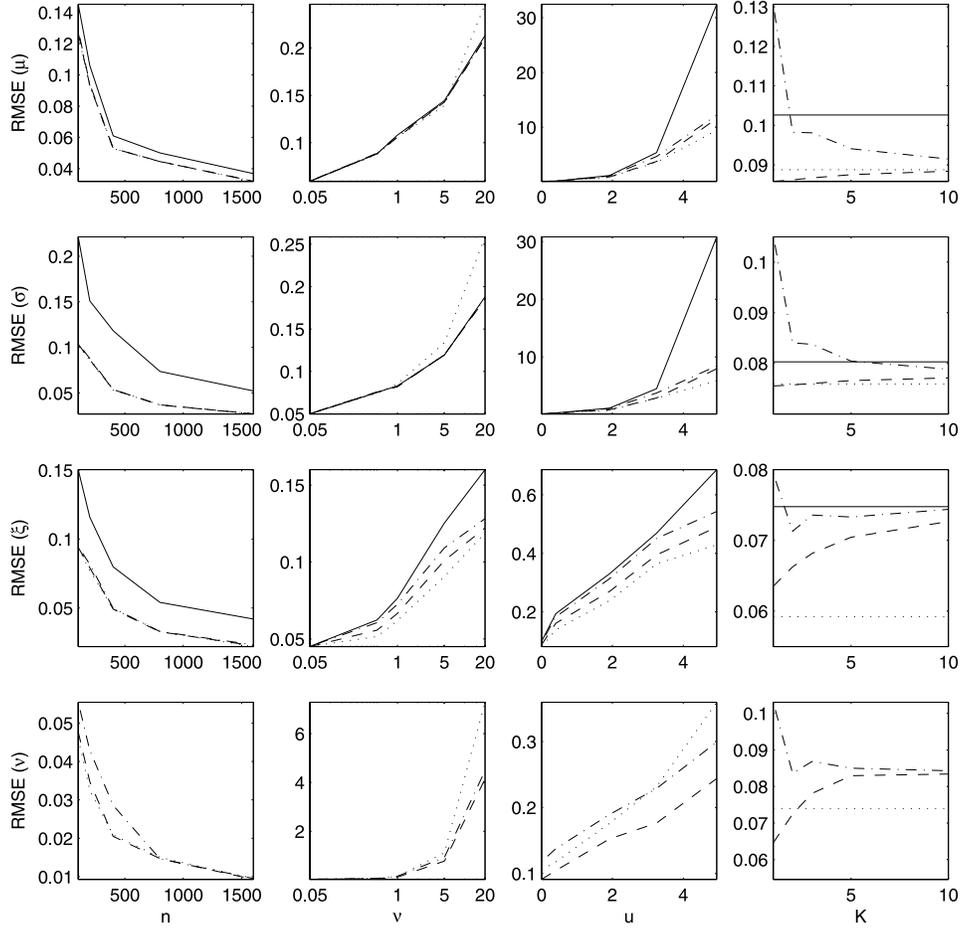}

\caption{RMSE ($y$-axis) as a function of sample
size $n$ (first column), dependence parameter $\nu$ (second column),
threshold $u$ (third column) and neighborhood size $K$ (last column).
Results obtained using $1000$ simulations of a Gaussian extreme value
process. Unless specified otherwise, the time sampling is regular and
the values $\mu=0$, $\sigma=1$, $\xi=0.3$, $\nu=0.5$, $u=-\infty$ and
$n=300$ are used. Solid line: MILE; dotted line: MMLE; dashed line: MPL$_1$E; dashed-dotted line:
MPL$_5$E. For the last column, the time step between successive
observations is drawn from a uniform distribution on the interval
$(0,2)$ and the dashed (resp., dashed-dotted) line corresponds to the
MPL$_K$E with the first (resp., second) weighting strategy
[see (\protect\ref{eqstrag1})] [resp., (\protect\ref{eqstrag2})].}\label{figRMSElong}
\end{figure}

Let us first focus on the case in which the time step between
successive observations is constant. According to the left plots of
Figure~\ref{figRMSElong}, all of the estimators seem to be
consistent. We also checked empirically that, when multiplied by $\sqrt
{n}$, the errors are almost constant, demonstrating that we retrieve
the usual speed of convergence. The $\mathrm{MILE}$ is clearly the least
efficient estimator, whereas the $\mathrm{MPL}_1\E$, $\mathrm{MPL}_5\E$ and $\mathrm{MMLE}$ produce similar results. A closer
look reveals, however, that the $\mathrm{MPL}_5\E$ is the least
accurate of these three estimators. The $\mathrm{MMLE}$ slightly
outperforms the $\mathrm{MPL}_1\E$ in estimating the scale
parameter $\sigma$ and the shape parameter~$\xi$, whereas the
$\mathrm{MPL}_1\E$ provides the best estimator of the dependence
parameter $\nu$. The second column of Figure~\ref{figRMSElong} shows
that the root-mean-square error (RMSE) of all the estimators decreases
with the dependence parameter $\nu$ and that it is more difficult to
obtain reliable estimates when the dependence between successive
observations is strong. The efficiency of the $\mathrm{MMLE}$ generally
deteriorates quicker when $\nu$ increases. It provides the worst
estimation of $\mu$, $\sigma$ and $\nu$ when $\nu\geq1$, but provides
the best estimation of $\xi$ for all the values of $\nu$ considered in
this experiment. These results indicate that the Markovian likelihood
may not be appropriate when the dependence is strong. The third column
of Figure~\ref{figRMSElong} depicts the behavior of the different
estimators when censoring occurs. As expected, they all worsen when the
threshold increases and the number of noncensored observations
decreases. It can be seen that the estimator which suffers most from
censoring is the $\mathrm{MILE}$, whereas the $\mathrm{MMLE}$ outperforms
both the $\mathrm{MPL}_1\E$ and $\mathrm{MPL}_5\E$ in
estimating the parameters of the marginal distribution. The $\mathrm
{MPL}_1\E$ again provides the best estimator of the dependence
parameter $\nu$.

%

The last column of Figure~\ref{figRMSElong} shows the influence of
the windows considered in defining the neighborhood, which are taken
into account in the pairwise likelihood functions. The Gaussian extreme
value process was simulated using an irregular time sampling (the time
lag between successive observations was drawn from a uniform
distribution on $[0,2]$) to allow comparison between the two strategies
discussed in Section~\ref{secCL}: the first involves the use of the
$K$ closest observations [see equation~(\ref{eqstrag1})], whereas the
second takes into account all observations falling within $K$ time
steps [see equation (\ref{eqstrag2})]. We found the first strategy
always to be the best. The evolution of RMSE with $K$ differs according
to the strategy used. It is increasing for the first strategy, meaning
that the best estimators are obtained with $K=1$, but generally
decreasing for the second, and the difference between the strategies
decreases when $K$ increases. Comparison with the $\mathrm{MILE}$ and
$\mathrm{MMLE}$ indicates again that even when the time sampling is
irregular the $\mathrm{MMLE}$ slightly outperforms the $\mathrm
{MPL}_1\E$ in estimating $\mu$, $\sigma$ and $\xi$ but the $\mathrm
{MPL}_1\E$ provides the best estimation of $\nu$.

The results given in this section suggest that the $\mathrm{MMLE}$ or
$\mathrm{MPL}_1\E$ are more favorable for practical applications
since their RMSE are generally the lowest. The two estimators have a
similar computational cost. Although the former may provide a slightly
better estimation when the dependence between successive observations
is small, it is clearly less efficient when the dependence is strong.
Finally, it seems reasonable to use the $\mathrm{MPL}_1\E$ in
practice and we will focus on this estimator in the sequel.

\section{Performance on classical time series models}\label{secsim}

The lack of data makes it generally difficult to validate models for
extreme values when facing real applications. In this section we
perform a simulation study to check whether the proposed methodology is
able to capture the extremal properties of several widely used time
series models. In Section~\ref{seclong} we simulate large samples to
obtain estimators with a low variance and to check whether the Gaussian
extreme value process provides an appropriate approximation of the
extremal behavior of the time series models under consideration. Then,
in Section~\ref{secshort} we simulate shorter time series to validate
the overall methodology in a more realistic context.

\subsection{Model validation}\label{seclong}

We focus on the following time series models:
\begin{itemize}
\item \textit{IID}: $\{X_t\}$ is an i.i.d. sequence of standard normal variables.
\item \textit{AR}(1): $\{X_t\}$ is a discrete time stationary process
which satisfies
\[
X_t = \alpha X_{t-1} + \sqrt{1-\alpha^2}
\varepsilon_t
\]
for all $t$, where $\alpha\in(-1,1)$ describes the dependence between
successive observations and $\{\varepsilon_t\}$ is an i.i.d. sequence of
standard normal variables. The marginal distribution of $\{X_t\}$ is
standard normal and the extremal index [see \citet{Coles2001}] is equal
to one (no clustering of extremes). We use the value $\alpha=0.2$.
\item \textit{logARMAX}(1): $\{X_t\}$ is a discrete time stationary
process which satisfies $X_t=\log(U_t)$, where $\{U_t\}$ is an ARMAX(1)
process defined as
\[
U_t=\max\bigl\{(1-\alpha) U_{t-1}, \alpha
\varepsilon_t \bigr\}
\]
for all $t$, where $\alpha\in(0,1)$ describes the dependence between
successive observations and $\{ \varepsilon_t\}$ is an i.i.d. sequence of
unit Fr\'echet variables. The logarithmic transformation is used to
avoid the numerical problems which occur when estimating quantities
related to heavy tail distributions by simulation. The marginal
distribution of $\{U_t\}$ is unit Fr\'echet, whereas the one of $\{X_t\}
$ is Gumbel. The extremal index is $\alpha$. We use the value $\alpha=0.2$.
\item \textit{OU} (Ornstein--Uhlenbeck): $\{X_t\}$ is a continuous time
stationary process which satisfies
\[
dX_t = - \alpha X_t \,dt + \sqrt{2\alpha}
\,dW_t,
\]
where $\alpha>0$ describes the dependence structure and $\{W_t\}$ is a
standard Brownian motion. The marginal distribution of $\{X_t\}$ is a
standard normal distribution, and we retrieve the AR(1) model when the
process is sampled at a regular time step. We use the value $\alpha
=0.05$, and the time lags between successive observations are drawn
from a uniform distribution on $[0,2]$.
\end{itemize}

%
\begin{figure}[b]

\includegraphics{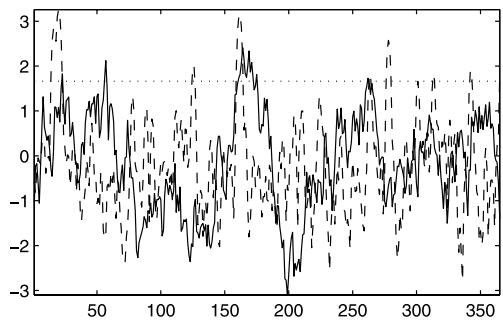}

\caption{Short samples of the AR(1) model (solid
line) and fitted Gaussian extreme value process (dashed line). The
horizontal dotted line is the threshold used for censoring.}\label{figARvsplot}
\end{figure}

%
\begin{figure}

\includegraphics{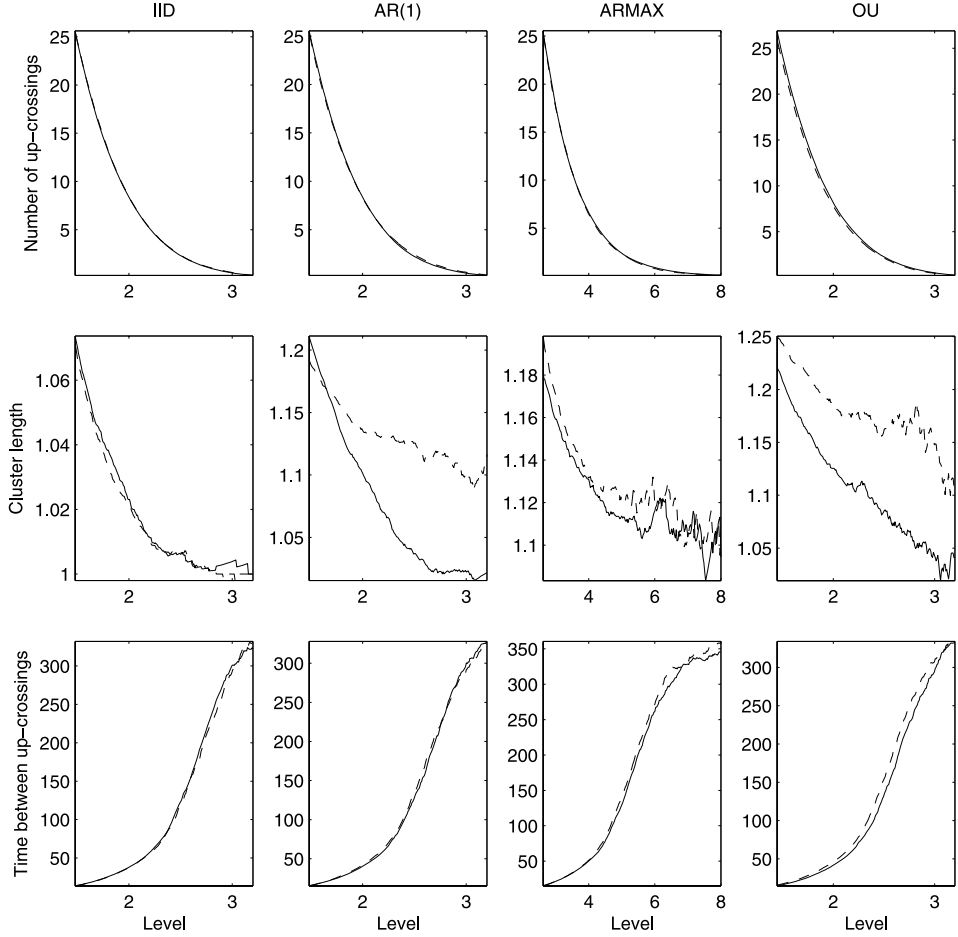}

\caption{Comparison of the extremal behavior of
classical processes (solid line) against the fitted Gaussian extreme
value process (dashed line). From top row: mean number of up-crossings
per year, mean length of clusters and mean time between consecutive
up-crossings as a function of the threshold ($x$-axis). Results obtained
by simulating 1000 years of each model (one observation per day).}\label{figvalidsim}
\end{figure}

For each of these models, we first generate a long realization
(equivalent to $1000$ years with one observation per day) and fit a
Gaussian extreme value process to the simulated sequence censored at
the $95\%$ quantile by computing the $\mathrm{MPL}_1\E$. We then
compare the following characteristics of the reference model and the
fitted censored Gaussian extreme value process:
\begin{itemize}
\item mean number of up-crossings during a given time period (one year)
as a function of the threshold;
\item mean length of the sojourns above a varying threshold; and
\item mean length of the sojourns below a varying threshold.
\end{itemize}
These quantities were selected because they summarize important
properties of the extremal behavior of the processes and are important
for practitioners. All quantities were computed using long simulations
of both the original time \mbox{series} model [IID, AR(1), logARMAX(1) or OU
generated using standard algorithms] and the fitted Gaussian extreme
value process. This is illustrated in Figure~\ref{figARvsplot} which
shows realizations of both the AR(1) model and the fitted Gaussian
extreme value process, whereas the second column of Figure~\ref{figvalidsim} permits a more systematic \mbox{comparison} of the extremal
behavior of both processes. The fitted model is able to reproduce both
the frequency of the up-crossings and the durations between successive
up-crossings, even for high thresholds, but it slightly overestimates
the mean length of the clusters above high thresholds. Indeed, for the
AR(1) model, the mean length of the clusters tends to one when the
threshold increases, as expected from theory (no clustering of
extremes), whereas the fitted Gaussian extreme value process exhibits
small extremal dependence and thus clustering of extremes. Using a
higher threshold for censoring before fitting the Gaussian extreme
value process helps to improve these results and to retrieve extremal
independence (not shown). According to the first column of Figure~\ref{figvalidsim}, the results are better for the IID model, which is a
particular case of the AR(1) model with no dependence between
successive observations. 
It is not surprising to obtain similar results for the OU (see the
last column of Figure~\ref{figvalidsim}) and AR(1) models since they
are equivalent when the time sampling is regular.

In contrast to the other models, the extreme values of the logARMAX(1)
process tend to cluster. Figure~\ref{figvalidsim} shows that the
fitted model seems to be able to reproduce both the frequency of the
up-crossings and the mean length of the clusters that tend to a limit
greater than one, as expected from theory. However, it seems to
overestimate the mean length of the sojourns below high thresholds,
although the erratic behavior of the curves suggests that the observed
differences may be due to sampling error.

These simulation results indicate that the Gaussian extreme value
process is able to capture important properties of the extremal
behavior of several common time series models, and similar results were
obtained for various other models we have tested (not reported).

%
\begin{table}[b]
\tabcolsep=0pt
\caption{Mean value of the estimated 100-year
return level with 90\% fluctuation intervals in parentheses. Simulation
results based on $200$ independent five-year synthetic sequences of
each model. A~declustering step was applied in the POT method [see
\citet{Coles2001}]}\label{reallifequantile}
\begin{tabular*}{\tablewidth}{@{\extracolsep{\fill}}@{}lcccc@{}}
\hline
\textbf{Method} & \textbf{IID} & \textbf{AR(1)} & \textbf{logARMAX(1)} & \textbf{OU} \\
\hline
True value & 4.03 & 4.03 & 8.90 & 3.79 \\[6pt]
POT & 4.18 (3.12, 6.02) & 4.16 (3.14, 5.68) &12.19 (5.31, 29.96) & 3.21 (2.31, 5.04)\\
$\mathrm{MMLE}$ & 3.89 (3.12, 5.15) & 3.90 (3.09, 4.95) & 18.77 (5.88, 36.96) & 3.62 ( 2.61, 5.05)\\
$\mathrm{CPL}_1\E$ & 3.84 (3.11, 5.17) & 3.76 (3.16, 4.72) & \phantom{0}9.52 (5.54, 17.09) &3.62 (2.61, 4.98)\\
\hline
\end{tabular*}
\end{table}

\subsection{Simulation results in a realistic context}\label{secshort}
In this section we validate our proposed methodology on shorter
synthetic time series that better correspond to the amount of data
typically available in environmental applications (i.e., a few years of
data). For each of the time series models introduced in the previous
section, we repeated the following numerical experiment $1000$ times:
\begin{itemize}
\item generate a five-year sequence (one observation per day) of the
reference time series model;
\item fit the censored Gaussian extreme value process to this sequence
after censoring at the $95\%$ quantile; and
\item compute the $100$-year return level for the fitted model which is
defined as the level at which the mean number of \textit{clusters}
above this level in a 100-year time period is equal to one. This return
level was chosen because it is generally the quantity of interest in
practical applications. It depends on both the marginal distribution
and the dependence structure of the process. It was computed by
simulating a long realization (1000 years) of the fitted model.
\end{itemize}

It can be seen from Table~\ref{reallifequantile} that the results
obtained with the fitted censored Gaussian extreme value process
clearly outperform those obtained with the usual POT method. The
$\mathrm{MMLE}$ and $\mathrm{CPL}_1\E$ produce similar results for the
three models with no extremal dependence [IID, AR(1) and OU], but those
obtained with the $\mathrm{CPL}_1\E$ are clearly superior to those
obtained with the $\mathrm{MMLE}$ in terms of accuracy and bias for the
logARMAX(1) model. The IID and AR(1) models have almost the same
100-year return levels, which is expected from theory since they have
the same marginal distribution and no extremal dependence. The lower
return period of the OU process, which also has the same marginal
distribution and no extremal dependence when observed at a regular time
step, may be due to the irregular time sampling. The extremal
dependence of the logARMAX(1) model leads to a higher return level than
the other models, but also to bigger bias and variance in the
estimators, which is in conformity with the simulation results reported
in Section~\ref{secsimu} (see Figure~\ref{figRMSElong}).

\section{Application to significant wave height}\label{secappli}

Significant wave height (Hs) is an important oceanographic parameter
that is directly related to the energy of a sea state. It was
originally defined as the mean height of the one-third highest waves,
and was thought to give about the same value as an experienced seaman's
eyeball estimate of wave height. With the development of instruments
producing more accurate measurements of sea surface elevation, Hs was
redefined as four times the standard deviation of the sea surface
elevation on a certain space--time domain where the sea state conditions
can be assumed to be stationary. The ratio of four was chosen to ensure
that the two definitions roughly coincide.

Offshore structures in particular must be designed to exceed a specific
level of reliability, which is typically expressed in terms of return
periods of Hs. The three sources of data (buoy, reanalysis and
satellite) that can be used to estimate the extremal behavior of Hs are
introduced in Section~\ref{secdata}. In Section~\ref{secsingle} we
focus on a specific location in which buoy data are available and
compare the results obtained with the three data sets. The buoy and
satellite data give similar results, whereas the reanalysis data lead
to significantly different results. Since the buoy is generally
considered to be the reference, it suggests the use of satellite data
when no buoy data are available at the location of interest. This is
further discussed in Section~\ref{secspat} which shows maps of Hs
return-periods in the North Atlantic based on satellite data.

\subsection{Hs data}\label{secdata}

The data used in this work come from the three sources briefly
described below:
\begin{itemize}
\item\textit{Reanalysis data}. The ERA-interim data set is a global
reanalysis carried out by the European Center for
Medium-Range Weather Forecasts (ECMWF). It can be freely downloaded and used for scientific
purposes.\footnote{\url{http://data.ecmwf.int/data/}.} In this work, we
consider 21 years of data, from 1989 until 2009.

\item\textit{Buoy data}. We focus on data from the buoy Brittany
(station 62,163\footnote{\url{http://www.ndbc.noaa.gov/station\_page.php?station=62163}.}), which is part of the UK Met Office monitoring
network. It is located at position (47.5 N, 8.5 W) and provides hourly
Hs data. In this work, we consider 10 years of data, from 1995 until
2005 (no data are available for 2000). Missing values represent about
7.7\% of the data and are generally associated with extreme events (as
breakdowns generally occur during storms), which is an important issue
when implementing block maxima or POT approaches.

\item\textit{Satellite data}.
The observations consist of Hs measured at discrete locations along
one-dimensional tracks from seven different satellite altimeters which
have been deployed progressively since 1991. The data set and related
information can be freely downloaded.\footnote{\url{ftp://ftp.ifremer.fr/ifremer/cersat/products/swath/altimeters/waves/}.}
In this study, we consider data from 1992 until 2007.
\end{itemize}

A typical example of data coverage over a 24-hour time window in the
North Atlantic Ocean can be seen in Figure~\ref{figdatadispo}. The
reanalysis data are available over a regular $1.5\times1.5$-degree
grid at synoptic times every six hours starting at midnight, in
contrast to the irregular space--time sampling provided by the satellite
altimeter. However, the reanalysis data tend to underestimate Hs
variability (see the next section) and they provide information only at
a synoptic scale, whereas satellite data give smaller-scale information
as well.

Buoys provide accurate information on sea-state conditions but are
sparsely distributed over the ocean, and there is generally no buoy at
the location of interest for a particular application. In such a
situation, it is important to be able to derive estimates of the
extremal behavior of Hs from the other sources of data introduced in
this section, which are available all over the oceans. These data sets
have been considered in numerous studies interested in the
distributional properties of Hs [see, e.g., \citet{Challenor1990},
\citet{Tournadre1990}, \citet{Caires2005}, \citet
{Memendez2008}, \citet{vinoth2011}].

\subsection{Single-site analysis}\label{secsingle}

In this section we focus on the location of the buoy Brittany and
illustrate how the methodology introduced in this work can be used to
estimate the extremal behavior of Hs using the three data sets
introduced in the previous section.

Hs data are nonstationary with an important seasonal component. A
classical approach to treating seasonality in meteorological
applications is to
block the data by month and fit a separate model each month, assuming
that the different realizations of the same month across years are
independent realizations
of a common stochastic process. We adopt this approach herein, focusing
on the month of December. Blocking the data by month leads, however, to
waste data and probably to the loss of important information on extreme
events. The development of nonstationary models which include seasonal
and inter-annual components to cover the whole time series would be a
valuable topic for future research.

%
\begin{figure}[b]

\includegraphics{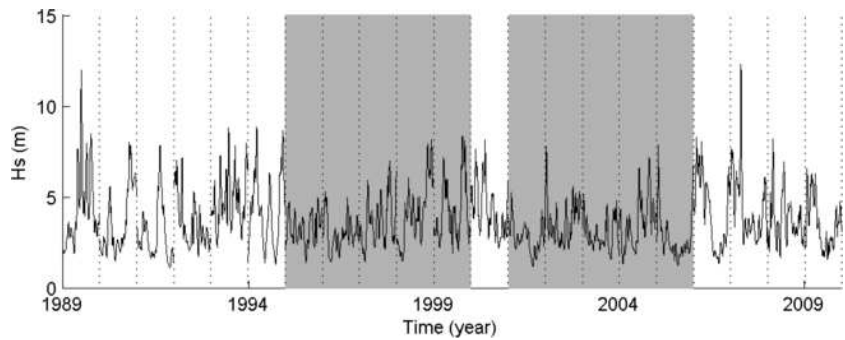}

\caption{Reanalysis data for 21 months of December. The
dotted vertical lines correspond to the beginning of the monthly blocks
(1st of December of each year) and the gray area corresponds to the
years in which both buoy and restricted reanalysis data are available.}\label{figERA}
\end{figure}

Reanalysis and satellite data provide observations that are not exactly
at the location of interest. Interpolation methods could be used, but
may smooth the data [see \citet{ailliot2011}]. We thus decided to consider:
\begin{itemize}
\item the reanalysis data available at location (48 N, 9 W), which is
the closest grid point to the buoy Brittany, and
\item the time series obtained by retaining all of the closest
observations to the buoy in the satellite tracks which intersect a
$3^\circ\times3^\circ$ box centered on the location of interest [see
\citet{vinoth2011}, \citet{wimmer2006}].\looseness=1
\end{itemize}

Figure~\ref{figERA} shows the resulting reanalysis time series, which
exhibits an important inter-annual variability particularly when we
look at the extreme values. There were two severe storms in which Hs
exceeded 12 meters in 1989 and 2007, whereas for the other years Hs was
always below 9 meters. These two storms exert a strong influence on the
results obtained from fitting a model to extreme events. To facilitate
comparison with the buoy data (which are unavailable for 1989 and
2007), we also consider a subset of reanalysis data for the years for
which buoy data are available (see Figure~\ref{figERA}). This data set
is referred to as ``restricted reanalysis data,'' whereas the full
reanalysis time series is named ``full reanalysis data.''

\begin{figure}[t]

\includegraphics{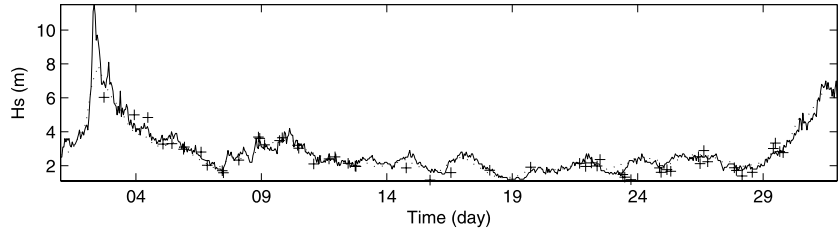}

\caption{Comparison of the three time series for
December 2005. Solid line: buoy data (location [47.5~N, 8.5~W]); dotted
line: reanalysis data (location [48~N, 9~W]); plus points: closest
satellite observation to the buoy from each satellite track within a
$3^\circ$ box.}\label{figERASat}
\end{figure}

\begin{figure}[b]

\includegraphics{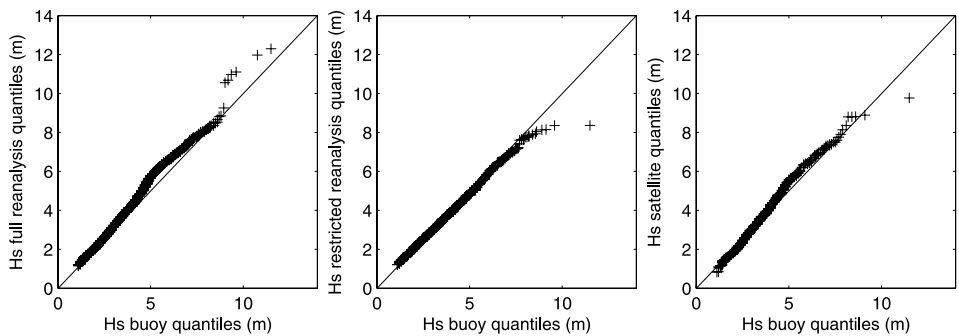}

\caption{QQ-plots of the empirical distribution
of the buoy ($x$-axis) against the empirical distribution of the full
reanalysis data ($y$-axis on the left panel), the reduced reanalysis data
($y$-axis on the middle panel) and the satellite data ($y$-axis on the
right panel).}\label{figqqplots}
\end{figure}

Figure~\ref{figERASat} shows all of the data available for December
2005. The agreement between the reanalysis and buoy time series is
generally good, although the reanalysis data tend to be smoother and to
exhibit lower extremes, as confirmed by the quantile--quantile (QQ)
plots in Figure~\ref{figqqplots}, which show the buoy data to have
higher quantiles than the restricted reanalysis data. However, this no
longer holds true when we compare the quantiles of the buoy and full
reanalysis data because of the significant inter-annual variability
(the months of December when the buoy data are available correspond to
years with generally low Hs). Finally, Figure~\ref{figqqplots} shows
good agreement between the empirical quantiles of the buoy and
satellite data, which suggests that the satellite data may constitute a
better source of information on high Hs than the reanalysis data.
Figure~\ref{figERASat} presents the complex temporal sampling of the
satellite data with clusters of several observations and long gaps with
no observation, which prevents the use of standard methods of extreme
value theory (i.e., block maxima and POT).

%
%

\begin{figure}[t]

\includegraphics{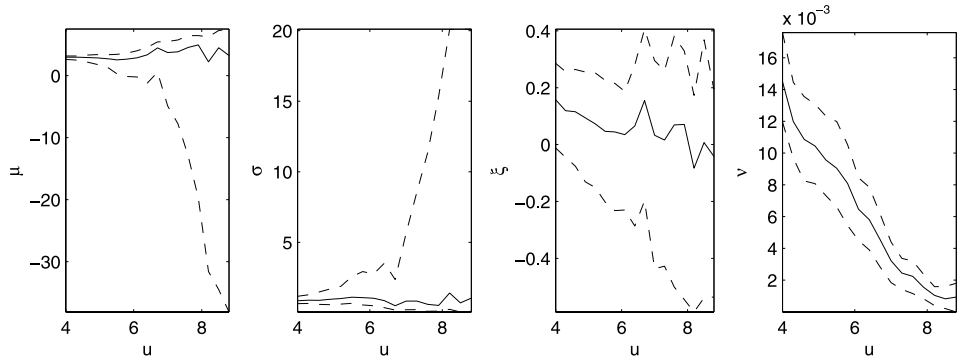}

\caption{Values of the MPL$_{1}$E computed on buoy data as
a function of threshold $u$ ($x$-axis).
From left to right: estimates of $\mu$, $\sigma$, $\xi$ and $\nu$. The
dashed lines correspond to 95\% confidence intervals computed using
parametric bootstrap (results based on $1000$ independent replications
of the fitted model simulated with the temporal sampling of the
original data).}\label{figparambouseuil}
\end{figure}

We first focus on buoy data since this data set is generally considered
to be the reference for the other data sets [see, e.g., \citet
{queffeulou2004}]. The first step is to choose a censoring threshold
$u$. This is a crucial step since $u$ must be high enough to justify
approximation by probabilistic models derived from extreme value theory
but not too high in order to keep enough observations to fit the model.
A~common tool for selecting an appropriate threshold is to fit the
model for various thresholds and choose the lowest one that ensures the
estimates are almost stable for any higher threshold value. Indeed,
from a theoretical point of view, if the fitted censored Gaussian
extreme value process is an appropriate model for describing the
behavior of the observed process above a threshold $u$, then it should
also be appropriate above a higher threshold. However, in practice, it
is generally difficult to come up with a decision using such diagnostic
plots. Figure~\ref{figparambouseuil} shows that the estimate of $\nu$
seems to stabilize only over the threshold $u=8$ meters, which has been
selected in the sequel. This threshold roughly corresponds to the 99\%
quantile of the marginal distribution.

%

%
\begin{table}[b]
\tabcolsep=0pt
\caption{Thresholds, parameter values and return levels
for the different data sets. 95\% confidence intervals computed using
parametric bootstrap are given in parentheses (results based on 1000
independent replications of the fitted models simulated with the
temporal samplings of the original data)}\label{resfit}
\begin{tabular*}{\tablewidth}{@{\extracolsep{\fill}}@{}ld{2.13}d{2.15}d{2.14}d{2.14}@{\hspace*{-2pt}}} \hline
& \multicolumn{1}{c}{\textbf{Full reanalysis}} & \multicolumn{1}{c}{\textbf{Restricted reanalysis}} & \multicolumn{1}{c}{\textbf{Buoy}} & \multicolumn{1}{c@{}}{\textbf{Satellite}} \\
\hline
\multicolumn{5}{@{}c@{}}{Threshold}\\[2pt]
$u$ & \multicolumn{1}{c}{\phantom{00}6} & \multicolumn{1}{c}{\phantom{0}6} & \multicolumn{1}{c}{\phantom{0}8} & \multicolumn{1}{c@{}}{\phantom{0}6}\\
Nb obs${}>u$ & \multicolumn{1}{c}{182} &\multicolumn{1}{c}{74} &\multicolumn{1}{c}{59} & \multicolumn{1}{c@{}}{48}
\\[6pt]
\multicolumn{5}{@{}c@{}}{Parameter values}\\[2pt]
$\mu$ &4.20\mbox{ }(3.20, 4.66) & 2.19\mbox{ }(-6.85, 3.80) &5.12\mbox{ }(-7.3, 6.67)& 3.80\mbox{ }(2.23, 4.47)\\
$\sigma$ &0.69\mbox{ }(0.41, 1.31)& 1.70\mbox{ }(0.81, 8.83) & 0.50\mbox{ }(0.11, 7.23)& 1.33\mbox{ }(0.86, 2.69)\\
$\xi$ &0.14\mbox{ }(-0.10, 0.3) &-0.17\mbox{ }(-0.6, -0.02) & 0.07\mbox{ }(-0.40, 0.32)& 0.01\mbox{ }(-0.25, 0.17)\\
$\nu$ & 0.13\mbox{ }(0.10, 0.17) & 0.12\mbox{ }(0.07, 0.18) &\multicolumn{1}{c}{1e--3\mbox{ }(8e--4, 2e--3)} & 0.05\mbox{ }(0.03, 0.08)
\\
\multicolumn{5}{@{}c@{}}{Return levels}\\[2pt]
$q_{10}$ & 10.1\mbox{ }(8.5, 12.3) & 8.2\mbox{ }(7.3, 9.1) & 10.4\mbox{ }(9.3, 12.2)& 12.1\mbox{ }(10.7, 14.6) \\
$q_{20}$ & 11.0\mbox{ }(9.0, 14.3) & 8.5\mbox{ }(7.5, 9.6) & 10.8\mbox{ }(9.4, 13.8)& 12.9\mbox{ }(11.1, 16.4) \\
$q_{50}$ & 12.5\mbox{ }(9.5, 17.9) & 8.8\mbox{ }(7.6, 10.1) & 11.4\mbox{ }(9.6, 16.0)& 14.0\mbox{ }(11.5, 19.1) \\
$q_{100}$ & 13.6\mbox{ }(9.9, 21.6) & 9.0\mbox{ }(7.7, 10.6) & 11.8\mbox{ }(9.7, 18.8)& 14.7\mbox{ }(11.7, 21.6) \\
\hline
\end{tabular*}
\end{table}

\begin{figure}

\includegraphics{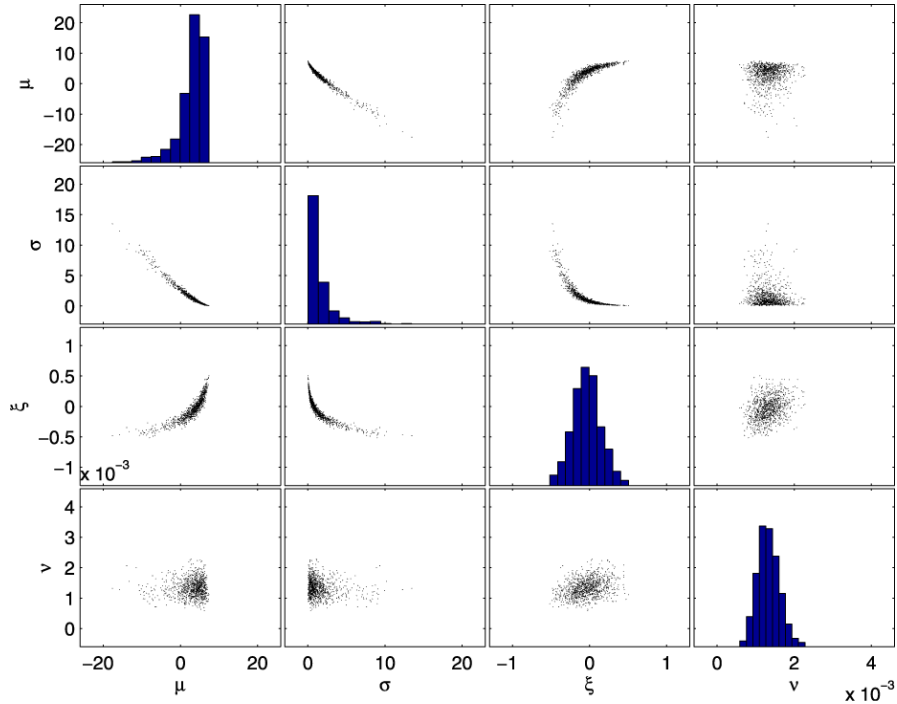}

\caption{Empirical distribution of the parameters
obtained using parametric bootstrap and the model fitted to buoy data.
From left to right: estimates of $\mu$, $\sigma$, $\xi$ and $\nu$. The
diagonal plots show histograms of the marginal distributions and the
other plots are scatter plots of the bivariate distributions (results
based on 1000 independent replications of the fitted model simulated
with the temporal sampling of the original data).}\label{fignuage}
\end{figure}

The parameter values are given in Table~\ref{resfit} together with 95\%
confidence intervals computed using parametric bootstrap.
The shape parameter $\xi$ is slightly negative (bounded tail), but the
sampling distribution shown in Figure~\ref{fignuage} indicates that
the difference from zero (Gumbel distribution) is not significant. This
is in agreement with previous studies [see, e.g., \citet{Caires2005}] in
which $\xi$~is often fixed to be equal to zero. More generally,
Figure~\ref{fignuage} shows the empirical distribution of the parameters
obtained using parametric bootstrap and the strong relations between
the estimates of the parameters $\mu$, $\sigma$ and $\xi$. High values
for the position parameter $\mu$ are generally associated with high
values for the shape parameter $\xi$, which is compensated for by low
values for the scale parameter $\sigma$. As a consequence, any error on
one parameter influences the values of the others, leading to wide
confidence intervals for $\mu$, $\sigma$ and $\xi$ (see Table~\ref
{resfit}). The parameter $\nu$ is less correlated with the other
parameters, although there is a noticeable positive correlation with
$\xi$. In applications, the end user is generally interested in return
levels, which are functions of the four parameters of the model, and,
thus, the errors made on individual parameters may compensate for one
another [see also \citet{Ribereau2011}].

%

\begin{figure}

\includegraphics{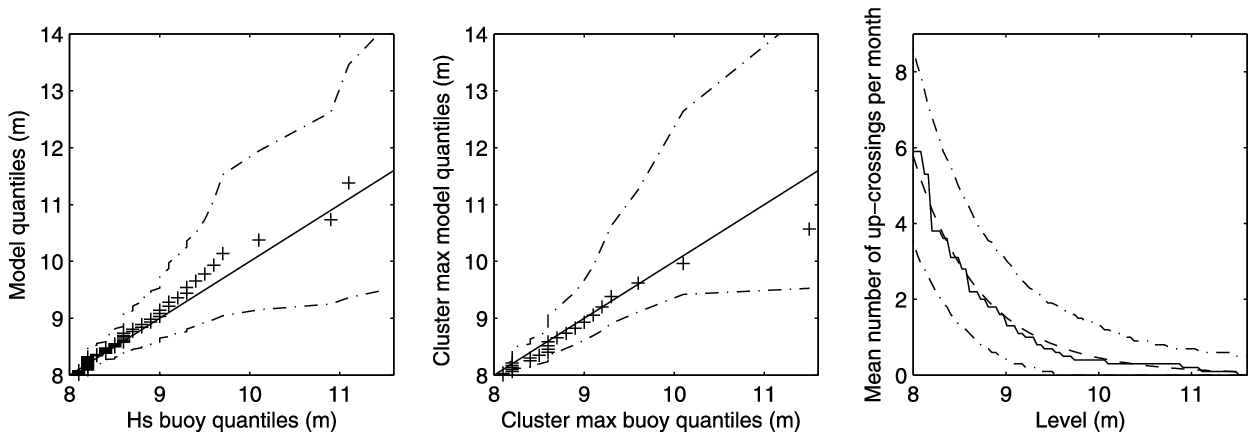}

\caption{Comparison of the extremal behavior of
the buoy data and fitted model. First panel: \mbox{QQ-}plot of the marginal
distribution of the buoy ($x$-axis) against the fitted model. Second
panel: \mbox{QQ-}plot of the cluster maxima of the buoy ($x$-axis) against the
fitted model ($y$-axis). Third panel:~mean number of up-crossings per
month as a function of the level ($x$-axis) for the buoy data (solid
line) and fitted model (dashed line). The dashed-dotted lines represent
95\% prediction intervals computed by simulation (results based on
1000 independent replications of the fitted model simulated with the
temporal sampling of the original data).}\label{figstatsbou}
\end{figure}

Figure~\ref{figstatsbou} compares various important characteristics of
the extremal behavior of the data with those of the fitted model. The
statistics computed from the data always lie in the 95\% prediction
intervals for the fitted model and, thus, the model seems to be able to
reproduce both the marginal distribution and the dynamics of the
observed time series above the selected threshold. This gives us
confidence in the results obtained when extrapolating the extremal
behavior of the data using the model. Using a lower threshold $u$ leads
to estimates with smaller variances (see Figure~\ref{figparambouseuil}), but the fitted models are no longer able to
reproduce the characteristics of the data above high levels (results
not shown).

%

The censored Gaussian extreme value process was also fitted to the
reanalysis and satellite data. For these data sets, we selected a lower
threshold $u=6$ meters. This choice was based on the same diagnostic
plots as those previously discussed for buoy data and practical
considerations (using the threshold $u=8$ meters would lead to
retaining fewer than $10$ observations above the threshold for the
satellite and restricted reanalysis data). The parameter values of the
models fitted on the buoy and satellite time series are broadly similar
(see Table~\ref{resfit}) but exhibit important differences with those
obtained on the two reanalysis data sets. In particular, the parameter
$\nu$, which describes the dynamics of the process, is higher for the
reanalysis data sets. This is in agreement with Figure~\ref{figERASat}, which shows that the reanalysis data tend to be smoother
than the buoy data. The value of $\xi$ is close to zero for the buoy
and satellite data, whereas it is significantly negative (bounded tail)
for the restricted reanalysis data and slightly positive for the full
reanalysis data, probably because of the influence of the two severe
storms in 1989 and 2007 (see Figure~\ref{figERA}). These results are
also in agreement with Figure~\ref{figqqplots}, which indicates that
the full reanalysis data set has a heavier tail than the buoy data set,
whereas the restricted reanalysis data set has a lighter tail.

\subsection{Spatial analysis}\label{secspat}

Similar results were obtained at the other locations for which buoy
data were available. The buoy and satellite data generally lead to the
identification of similar models, whereas the reanalysis data identify
more extremal dependence and longer storms. If we believe the buoy data
to be a good reference, then these results suggest that satellites may
provide more accurate information on the extremal behavior of Hs than
reanalysis data. In this context, the proposed methodology can be an
efficient tool for estimating the extremal properties of Hs at any
ocean location for which satellite data are available.

\begin{figure}

\includegraphics{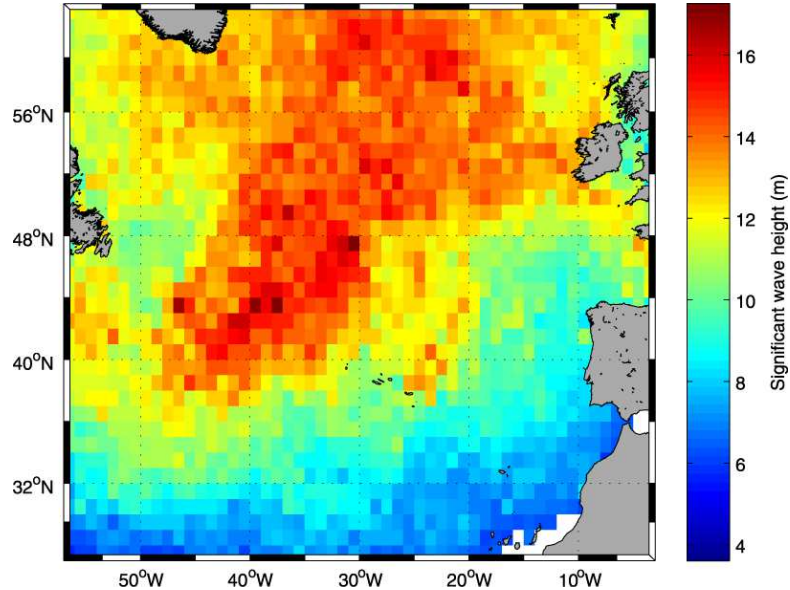}

\caption{Map of 20-year return levels in the North
Atlantic Ocean. A $3^\circ$-wide moving window was used to build the
time series at the different locations by retaining the value closest
to the center for each satellite track crossing the window. A variable
threshold corresponding to the 95\% quantile was used.}\label{figmapssat}
\end{figure}

This is illustrated in Figure~\ref{figmapssat}, which shows $20$-year
return levels in the North Atlantic computed using satellite data. To
reduce the variability of the estimates, we have fixed the tail
parameter $\xi$ to be equal to zero, which is a common assumption when
fitting extreme value models to Hs data [see, e.g., \citet{Caires2005}]
and is in agreement with our previous single-site analysis. The map in
Figure~\ref{figmapssat} exhibits a clear spatial structure, coherent
with that obtained using another method in \citet{wimmer2006}. Further
improvements may be obtained by constraining the parameters so that
they vary smoothly in space [see, e.g., \citet{Cooley2007},
\citet{reich2012hierarchical}], although such a sophisticated
development is
beyond the scope of this study.

%

\section{Conclusion}\label{secconclu}
In this paper we propose an original method for analyzing the extremal
behavior of univariate time series. Our approach is motivated by the
need to analyze environmental time series with missing values or
irregular sampling. Tests performed on classical time series models
indicate that the proposed method also performs well on time series
with regular sampling compared to other methods proposed in the
literature. The parameters are estimated using a composite likelihood
method, and both theory and simulations indicate that doing so leads to
consistent estimates. The results obtained on Hs data indicate that the
proposed methodology can be used to estimate the extremal behavior of
Hs from satellite data and produce an accurate climatology of extreme
Hs all over the ocean.

We believe our methodology to be sufficiently flexible to build
extensions useful for a range of applications. For example, it could
deal with other max-stable processes, include nonstationary components
or be extended to a space--time model. These possible extensions will be
the subject of future research.

\section*{Acknowledgments}
The authors are indebted to the Laboratoire d'Oc\'{e}ano\-graphie
Spatiale, IFREMER and to the ECMWF for providing the data used in this
study. We are also grateful to the anonymous reviewer, Associate Editor
and Editor for their constructive comments and suggestions, which have
led to significant improvements to the paper.

\begin{supplement}[id=suppA]
\stitle{Supplementary material: Proof of the consistency of the MPL\tsub{1}E estimates.}
\slink[doi]{10.1214/13-AOAS711SUPP}
\sdatatype{.pdf}
\sfilename{AOAS711\_supp.pdf}
\sdescription{In the attached supplemental material [\citet
{Raillard2013}], we prove the consistency of the $\mathrm{MPL}_1\E$
estimator in an idealized situation with no censoring and known
marginal distributions.}
\end{supplement}
%


%

\printaddresses

\end{document}